\documentclass[a4paper,12pt]{article}

\usepackage{typearea}

\usepackage{t1enc}
\usepackage{mathrsfs}
\usepackage{graphicx}
\usepackage{multirow}
\usepackage{hyperref}
\usepackage{bbm}
\usepackage{amsmath}
\usepackage{arydshln}
\usepackage{tikz}
\usepackage{authblk}
\usepackage[hang,small,bf]{caption}

\newcommand{\qtilde}{\tilde{q}}

\usepackage{tikz}
\usetikzlibrary{shadows}
\usetikzlibrary{arrows,shapes,positioning}
\usetikzlibrary{decorations.markings}
\usetikzlibrary{decorations.pathmorphing}
\usetikzlibrary{decorations.pathreplacing}
\tikzstyle arrowstyle=[scale=1]
\tikzstyle directed=[postaction={decorate,decoration={markings, mark=at position .65 with {\arrow[arrowstyle]{stealth}}}}]
\tikzstyle end directed=[postaction={decorate,decoration={markings, mark=at position 1 with {\arrow[arrowstyle]{stealth}}}}]
\tikzstyle reverse directed=[postaction={decorate,decoration={markings, mark=at position .65 with {\arrowreversed[arrowstyle]{stealth};}}}]
\tikzstyle{ann} = [fill=white,font=\footnotesize,inner sep=1pt]

\title{
  \flushright
  {\footnotesize SFB/CPP-14-124, DESY-15-059, CP3-Origins-2015-013 DNRF90, DIAS-2015-13}
  \vspace{5mm}\\
  \centering
  {\bf A Lattice Calculation of Parton Distributions}
  }

\author[a,b]{\normalsize Constantia Alexandrou}
\author[c,d,e]{Krzysztof Cichy}
\author[f]{Vincent Drach}
\author[c,g]{Elena Garcia-Ramos}
\author[a]{Kyriakos Hadjiyiannakou}
\author[c]{Karl Jansen}
\author[c]{Fernanda Steffens}
\author[c]{Christian Wiese}

\affil[a]{\footnotesize Department of Physics, University of Cyprus, P.O. Box 20537, 1678 Nicosia, Cyprus}
\affil[b]{The Cyprus Institute, 20 Kavafi Street, Nicosia 2121, Cyprus}
\affil[c]{John von Neumann Institute for Computing (NIC), DESY, Platanenallee 6, 15738 Zeuthen, Germany}
\affil[d]{Faculty of Physics, Adam Mickiewicz University, Umultowska 85, 61-614 Pozna\'{n}, Poland}
\affil[e]{Goethe-Universit\"at, Institut f\"ur Theoretische Physik, Max-von-Laue-Strasse 1, \newline 60438 Frankfurt a.M., Germany} 
\affil[f]{CP$^3$-Origins \& the DIAS, University of Southern Denmark, Campusvej 55, 5230 Odense M, Denmark}
\affil[g]{Humboldt-Universit\"at zu Berlin, Institut f\"ur Physik, Newtonstrasse 15, 12489 Berlin, Germany }

\date{\vspace{-15mm}}

\begin{document}
\maketitle
\begin{abstract}
We report on our exploratory study for the direct evaluation of the parton distribution functions
from lattice QCD, based on a recently proposed new approach. 
We present encouraging results using $N_f=2+1+1$ twisted mass fermions 
with a pion mass of about $370$\,MeV.
The focus of this work is a detailed description of the computation, including the 
lattice calculation, the matching to an infinite momentum and the nucleon mass correction.
In addition, we test the effect of gauge link smearing in the operator to estimate the influence of the
Wilson line renormalization, which is yet to be done.
\end{abstract}

\section{Introduction}

Parton distribution functions (PDFs) describe the structure of hadrons by providing 
information on the momentum, angular momentum and spin of quarks and gluons in a hadron. 
Ideally, PDFs would be directly predicted by quantum chromodynamics (QCD). 
Confronted with results from deep inelastic scattering experiments,
this would lead 
to a most stringent test of QCD and a deep theoretical understanding 
of the interaction between quarks and gluons. Naturally, lattice 
QCD methods, which can cover a broad energy range from the perturbative to 
the non-perturbative regimes, would be most suitable to compute the PDFs. 
However, such a calculation requires 
light-cone
dynamics and going to short, or even zero distance on the Euclidean space-time lattice 
is not possible. 

Nevertheless, through the 
operator product expansion, moments of the PDFs
can be expressed 
in terms of matrix elements of local operators, which are accessible to lattice QCD calculations. 
In fact, lattice QCD calculations of the PDF's moments have been very successfully
carried out with results emerging now directly at the physical value
of the pion mass, 
see the recent reviews of Refs.\,\cite{Alexandrou:2013cda,Alexandrou:2014yha,Syritsyn:2014saa,Constantinou:2014tga}.

Despite the enormous activity of computing such moments in lattice QCD, it 
would still be highly desirable to have information on the PDFs themselves. 
A reconstruction of the PDFs from their moments seems unfeasible on the 
lattice, since higher moments show a very bad signal-to-noise ratio and are 
very hard to compute. A solution to this problem might be 
the proposal in Ref.\,\cite{Ji:2013dva}, which suggests that by computing a
parton quasidistribution function, a quantity accessible to lattice 
computations, contact to the required physical PDFs can be established 
through a matching procedure. Such a matching has already been worked out 
in 1-loop perturbation theory \cite{Xiong:2013bka} and a first 
test of the approach has been carried out in Ref.\,\cite{Lin:2014zya} using staggered
fermions. 

Here, we will use a different quark discretization on the lattice, namely 
twisted mass fermions at maximal twist \cite{Frezzotti:2003ni}, to conduct
an exploratory study of the proposal in Ref.\,\cite{Ji:2013dva}.
This lattice formulation of QCD has the advantage that all physical 
quantities scale with a rate of $O(a^2)$ towards the continuum limit, 
and it avoids the operator improvement necessary in other lattice QCD
formulations, easing thus considerably the computations. Twisted mass lattice QCD 
calculations for baryons have already been carried out successfully  
for the baryon spectrum 
\cite{Alexandrou:2008tn,Alexandrou:2009qu,Alexandrou:2014sha}, 
for form factors and moments of PDFs 
\cite{Alexandrou:2010hf,Alexandrou:2011nr,Alexandrou:2011db,Dinter:2011sg,Alexandrou:2013joa}
and also for 
disconnected contributions to nucleon observables 
\cite{Alexandrou:2012zz,Abdel-Rehim:2013wlz,Alexandrou:2013wca}. 

As stated above, our work here focuses on exploring the potential of the approach 
in Ref.\,\cite{Ji:2013dva}. To this end, we concentrate on one 
ensemble of maximally twisted mass fermions at a lattice spacing 
of about 0.08\,fm and a pion mass of about 370\,MeV. 
 
In our calculations, we obtain results for a boosting 
nucleon frame, using the three lowest lattice momenta,
$2\pi/L, 4\pi/L$ and $6\pi/L$. 
Larger momenta show a signal-to-noise ratio that is
too poor to extract any meaningful result. 
We compute the real and the imaginary parts of the relevant matrix elements and 
find that the imaginary part is very important to give an asymmetry 
between the quark and anti-quark distributions, a highly non-trivial 
result of our calculation. 
In addition, we apply different levels of gauge link smearing in the operator. This smearing 
procedure has two effects. First, higher smearing levels reveal the asymmetry 
between quark and anti-quark distribution much clearer. Second, different smearing 
levels correspond to different renormalization properties of the matrix elements 
considered. Thus, comparing results from different smearing levels can give a hint
about the importance of renormalization, depending on the size 
of effects from smearing. 
We will finally use the matching condition to relate the quasidistribution 
to the real PDF and also apply nucleon target mass correction.

It needs to be stressed that the work presented here is only a 
very first step to understand the potential 
of the approach of Ref.\,\cite{Ji:2013dva}. 
It would be very important to look at larger momenta than used here 
to test that the perturbative matching works.
Using a hypothetical mixed momentum setup (described below), 
we illustrate that a satisfactory agreement with phenomenological 
investigations could be obtained if larger momenta were available.
We are planning to employ larger momenta in our next calculations 
by increasing our statistics by about an order 
of magnitude. 

\section{Theoretical setup}

A method to calculate 
quark distributions directly on a Euclidean lattice has recently been proposed \cite{Ji:2013dva}.
If successful, this method can greatly improve our comprehension of 
the structure of hadrons, as well as being the first {\it ab initio} 
QCD calculation of the Bjorken-$x$ dependence of the quark distributions. 
The key observation in this proposal is that from the general form of the matrix 
element of a twist-2 operator between a nucleon state with momentum $P=(P_0, 0, 0, P_3)$, 
\begin{equation}
\langle P| O^{\mu_1 \mu_2 ... \mu_n} |P \rangle = 2 a_n^{(0)} \Pi^{\mu_1 \mu_2 \dotsc \mu_n},
\label{eq1}
\end{equation}
a suitable choice of the indices $\mu_1, \mu_2, \dotsc, \mu_n$ makes sure 
that the corresponding distribution is a purely spatial correlation. 
In Eq.\,(\ref{eq1}), $a_n^{(0)}$ are the moments of the 
quark distributions and $ \Pi^{\mu_1 \mu_2 ... \mu_n}$ 
is a symmetric rank - $n$ tensor which can be 
formed with the target momentum $P$, as first calculated 
by Georgi and Politzer \cite{Georgi:1976ve}. Let $n=2k$, then
\begin{equation}
\Pi^{\mu_1 \mu_2 ... \mu_n} = \sum_{j=0}^k (-1)^j \frac{(2k - j)!}{2^j (2k)!}\{g ... g P ... P\}_{k,j} (P^2)^j,
\label{eq2}
\end{equation}
where the term $\{g ... g P ... P\}_{k,j}$ means a symmetric sum of $(2k)!/2^j j! (2k-2j)!$ distinct products of the form
$g^{\mu_1 \mu_2} ... g^{\mu_{2j-1} \mu_{2j}} P^{\mu_{2j+1}} ... P^{\mu_{2k}}$. Thus, setting $\mu_1 = \mu_2 = ... = \mu_{2k} = 3$, one gets
\begin{equation}
\Pi^{3 ... 3} = \sum_{j=0}^k (-1)^j \frac{(2k - j)!}{2^j (2k)!} \frac{(2k)!}{2^j j! (2k-2j)!} (-1)^j (P_3^2)^{k-j} (M^2)^j
\end{equation}
or
\begin{equation}
\langle P| O^{3 ... 3} |P \rangle = 2 \tilde{a}_{2k}^{(0)} (P_3)^{2k} \sum_{j=0}^{k} \mu^j \left( \begin{array}{c} 
2k-j \\
j \end{array} \right) \equiv 2 \tilde{a}_{2k} (P_3)^{2k},
\label{eq3}
\end{equation}
with $\mu = M^2/4(P_3)^2$ and $M$ the nucleon mass. 
Here, we have introduced $\tilde{a}_{2k}$ as the matrix elements 
of the operator without subtracting the corrections in the nucleon mass. 
In the end, we want the matrix elements $ \tilde{a}_{2k}^{(0)}$, 
which can be related to the usual moments of the 
quark distributions in the Infinite Momentum Frame (IMF). 
For now, we define
\begin{equation}
\tilde{a}_{n} (\Lambda, P_3) = \int_{-\infty}^{+\infty} x^{n-1} \tilde{q} (x,\Lambda,P_3) dx,
\label{eq4}
\end{equation}
and apply the inverse Mellin transformation to Eq.\,(\ref{eq3}) to obtain:
\begin{equation}
\label{eq5}
\qtilde(x, \Lambda, P_3) = \int_{-\infty}^\infty \frac{dz}{4\pi} e^{-izk_3} \langle P |
\bar{\psi}(0,z)\gamma^3 W(z) \psi(0,0) |P\rangle ,
\end{equation} 
where $\Lambda$ is the UV regulator, $k_3 = x P_3$ 
is the quark momentum in the $z$-direction, 
and $W(z) = e^{-ig \int_0^z dz^{'} A_3(z^{'})}$ is the Wilson line introduced 
to make the quark distribution gauge invariant. 
Eq.\,(\ref{eq5}) is called a quasidistribution because
it does not have the usual properties of a quark distribution. 
Most notably, the momentum fraction $x$ can be bigger than 1 or 
smaller than 0. Also, as discussed in Ref.\,\cite{Xiong:2013bka}, 
the calculation of the leading UV divergences to the quasidistributions in
perturbation theory are done 
keeping $P_3$ fixed while taking $\Lambda \rightarrow \infty$. 
This is in contrast to the case of the usual parton distributions, 
where one takes the limit $P_3 \rightarrow \infty$ first, that is, one first goes to the
IMF.  
The dependence on the UV regulator, $\Lambda$, will be 
translated, in the end, into a renormalization scale $\mu_R$ 
when relating the quasidistribution at finite $P_3$ to its 
counterpart at infinite $P_3$. For now, as we still do not have 
 a renormalization procedure for the operator and the 
coupling, we freely identify the UV regulator in the 
perturbative corrections in the case of the IMF with $\mu_R$, 
the renormalization scale, while keeping it as $\Lambda$ 
for the case of the quasidistributions.

To relate the quasidistributions to the usual quark distributions, 
one uses the fact that the infrared region of the 
distributions is untouched when going from a finite 
to an infinite momentum\footnote{An effective field theory 
approach to extract the parton distributions from the lattice
observables, using a systematic expansion in inverse 
powers of the nucleon momentum, was proposed in Ref.\,\cite{Ji:2014gla}.}. 
In other words, if $q(x,\mu_R)$ is the usual distribution 
defined though light-cone correlations, then one should have:
\begin{equation}
q(x,\mu_R) = q_{bare}(x)\left\{1 + \frac{\alpha_s}{2\pi} Z_F(\mu_R) \right\} + \frac{\alpha_s}{2\pi}\int_x^1 q^{(1)} (x/y,\mu_R) q_{bare}(y) \frac{dy}{y} + \mathcal{O}(\alpha_s^2),
\label{eq6}
\end{equation}
\begin{equation}
\tilde{q} (x,\Lambda,P_3) = q_{bare}(x)\left\{1 + \frac{\alpha_s}{2\pi} \tilde{Z_F}(\Lambda,P_3) \right\} + \frac{\alpha_s}{2\pi}\int_{x/x_c}^1 \tilde{q}^{(1)} (x/y,\Lambda,P_3 ) q_{bare}(y) \frac{dy}{y} + \mathcal{O}(\alpha_s^2),
\label{eq7}
\end{equation}
where $q_{bare}$ is the bare distribution, 
$Z_F$ and $\tilde{Z_F}$ are the wave function corrections 
and $q^{(1)}$ and $\tilde{q}^{(1)}$ are the
vertex corrections. Notice that the lower limit 
of integration in Eq.\,(\ref{eq7}) is $x/x_c$, where $x_c \sim \Lambda/P_3$
is the largest possible value of $x$ which renders the vertex and wave function
corrections to the quasidistributions meaningful. 
Opposite to the infinite momentum calculation, 
at finite $P_3$ the terms $\tilde{Z_F}$ and
$\tilde{q}^{(1)}$ do not vanish for $x > 1$, and thus this region has to 
be included, with the cut being made at $x>1$, but below $x_c$. On the other hand, because 
$ q^{(1)} (x,\mu_R ) = 0 $ for $x \geq 1$, 
the integration range in Eq.\,(\ref{eq6}) can
be extended down to $x/x_c$ as well.
 
Lattice simulations can be used to calculate the 
left hand side of Eq.\,(\ref{eq7}) through Eq.\,(\ref{eq5}). 
Ideally, one would use perturbation theory to an arbitrary order to calculate the right 
hand side of Eqs.\,(\ref{eq6}) and (\ref{eq7}) 
to extract the quark distribution. Currently, however, 
the self-energy and vertex corrections are known to $ \mathcal{O}(\alpha_s)$ only and
for the non-singlet case \cite{Xiong:2013bka}. 
With this in mind, Eqs.\,(\ref{eq6}) and (\ref{eq7}) can be combined to give
\begin{eqnarray}
\tilde{q} (x,\Lambda,P_3) &=& q(x, \mu_R) + \frac{\alpha_s}{2\pi} q(x,\mu_R) \left\{\tilde{Z_F}(\Lambda,P_3)-Z_F(\mu_R) \right\} \nonumber\\*
& & + \frac{\alpha_s}{2\pi}\int_{x/x_c}^1 \left(\tilde{q}^{(1)} (x/y,\Lambda,P_3 )-q^{(1)} (x/y,\mu_R )\right) q(y, \mu_R) \frac{dy}{y} + \mathcal{O}(\alpha_s^2) ,
\label{eq8}
\end{eqnarray}
and this is equivalent to Eq.\,(13) of Ref.\,\cite{Xiong:2013bka} 
if we consider quarks only. Notice that the quark number is conserved in the above expression, as long
as the integrals $\tilde{Z_F}(\Lambda,P_3)$, listed in the Appendix, have also a cut in $x_c$. We define 
$\delta Z_F^{(1)} (\mu_R/P_3, \Lambda/P_3) =\tilde{Z_F}(\Lambda,P_3)-Z_F(\mu_R)$ and
$Z^{(1)} (\xi,\mu_R/P_3, \Lambda/P_3) = \tilde{q}^{(1)} (\xi,\Lambda,P_3 )- q^{(1)} (\xi,\mu_R )$. 
One can include antiquarks using the crossing relation 
$\bar{q}(x) = - q(-x)$, and then rewrite Eq.\,(\ref{eq8}) as
\begin{eqnarray}
q(x, \mu_R) &=& \tilde{q} (x,\Lambda,P_3) - \frac{\alpha_s}{2\pi} \tilde{q} (x,\Lambda,P_3) \delta Z_F^{(1)} \left( \frac{\mu_R}{P_3}, \frac{\Lambda}{P_3}\right) \nonumber\\*
& & - \frac{\alpha_s}{2\pi}\int_{-1}^1
 Z^{(1)} \left( \frac{x}{y}, \frac{\mu_R}{P_3}, \frac{\Lambda}{P_3} \right) \tilde{q} (y,\Lambda,P_3) \frac{dy}{|y|} + \mathcal{O}(\alpha_s^2) ,
\label{eq9}
\end{eqnarray}
where we have solved the system for $q(x, \mu_R)$. 
The form of Eq.\,(\ref{eq9}) that we implement in the 
actual calculations is detailed in the Appendix.

Eq.\,(\ref{eq9}) can be improved by calculating the 
corrections in $M/P_3$ to an arbitrary order. 
As before, we write 
$\tilde{a}^{(0)}_{n} = \int_{-\infty}^{+\infty} x^{n-1} \tilde{q}^{(0)}(x,P_z) dx$ 
and use this definition, together with Eq.\,(\ref{eq4}), 
to Mellin invert Eq.\,(\ref{eq3}). After some manipulation ({\it cf.} \cite{Steffens:2012jx}), the result is:
\begin{equation}
\tilde{q}(x,P_z) = \frac{1}{1 + \mu \xi^2} \tilde{q}^{(0)}(\xi, P_z),
\label{eq11}
\end{equation}
where $\xi = \frac{2x}{1+\sqrt{1+4\mu x^2}}$ is the Nachtmann variable. 
The matching and the nucleon mass corrections are interchangeable.

\section{Lattice calculation}

In this section, we will describe our lattice setup 
and our lattice computations.

\subsection{Matrix elements on the lattice}

On the lattice, the bare matrix elements $h(P_3,z)$,
which appear in Eq.\,(\ref{eq5}), can be computed as 
\begin{align}
h(P_3,z) =\left \langle P \vert \overline{\psi}(z)\gamma_{3} W_{3}(z,0)\psi(0)\vert P \right\rangle,
\end{align}
with the Euclidean momentum $P=(0, 0, P_3, P_4)$ and $z=(0,0,z,0)$. 
Due to the (spatial) rotational symmetry on the lattice
the computation can be straightforwardly applied to the other 
spatial directions. Our final result will then be an
average over these three directions.

\begin{figure}
  \centering
  \includegraphics{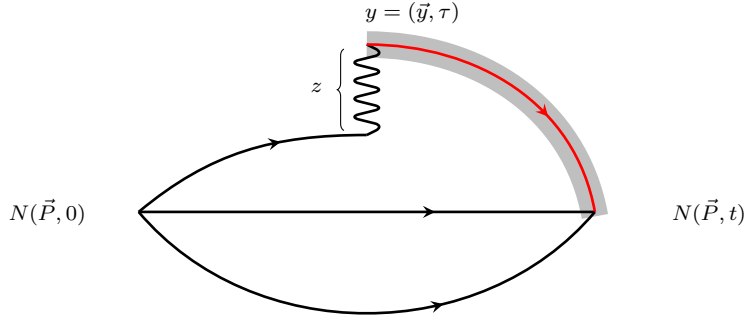}
  \caption{\label{FIG_DIA}Schematic picture of a possible Wick contraction of the quark fields in the three-point function.}
\end{figure}

The required matrix elements can be obtained from the ratio of suitable two-
and three-point functions. The three-point function
is constructed with the use of nucleon interpolating fields 
and a local operator:
\begin{align}
C^{\text{3pt}}(t,\tau,0) = \left \langle N_{\alpha}(\vec{P},t) \mathcal O(\tau) \overline{N}_{\alpha}(\vec{P},0)\right \rangle,
\end{align} 
where $\langle ... \rangle$ denotes the average over a sufficient number
of gauge field configurations. A nucleon field boosted with a three-momentum can be 
defined via a Fourier transformation of quark fields in position space:
\begin{align}
N_{\alpha}(\vec{P},t) = \Gamma_{\alpha\beta} \sum_{\vec{x}}\text{e}^{i \vec{P} \vec{x}}\epsilon^{abc}u_{\beta}^a(x)\left( {d^b}^T(x)\mathcal C \gamma_5 u^c(x)\right),
\end{align}
where $\mathcal C = i\gamma_0\gamma_2$ and $\Gamma_{\alpha\beta}$ is a suitable parity projector. 
Here, we will use the parity plus projector $\Gamma = \frac{1+\gamma_4}{2}$.
The matrix element at vanishing momentum transfer ($Q^2=0$)
can be obtained by choosing the following operator:
\begin{align}
\mathcal O(z, \tau, Q^2=0) = \sum_{\vec{y}}\overline{\psi}(y + z)\gamma_3 W_3(y+z,y)\psi(y),
\label{eq:operator}
\end{align}
with $y=(\vec{y},\tau)$. After Wick contracting the quark
fields, the three-point function can be expressed in terms
of quark propagators, see Fig.\,\ref{FIG_DIA} for a schematic picture of such a 
contraction.

We can extract the matrix element from a ratio of the above given three- and 
two-point functions:
\begin{align}
\frac{C^{\text{3pt}}(t,\tau,0;\vec{P})}{C^{\text{2pt}}(t,0;\vec{P})}\stackrel{0\ll \tau\ll t}{=}\frac{-iP_3}{E}h(P_3,z),
\end{align}
where $E=\sqrt{(P_3)^2+M^2}$ is the total energy of the
nucleon and $C^{\text{2pt}}$ is
the nucleon two-point function, which is constructed from the nucleon
interpolating fields,
$C^{\text{2pt}}(t,0;\vec{P})=\langle N_{\alpha}(\vec{P},t) \overline{N}_{\alpha}(\vec{P},0)\rangle$. 
For the operator, we will consider the non-singlet, iso-vector quark combination, symbolically represented
by $u-d$, which avoids disconnected contributions.   

When computing the three-point function, there is a freedom on how to treat the
propagator connecting the sink position with the operator insertion point
(highlighted in Fig.\,\ref{FIG_DIA}).
Due to momentum projection, there is a spatial sum on both ends of the propagator, which would 
naively require an all-to-all propagator. However, such a computation 
would need $V=L^3\times T$ sets of inversions.

Here, we have tested 
two different methods to calculate the propagator.
The first is the sequential method, which is exact. However, it requires 
the sink position and momentum to be fixed.
As a second choice, we have used 
a stochastic method, where we use sources that contain
$Z^4$ noise on one single timeslice ({\it cf.} Ref.\,\cite{Alexandrou:2013xon}). 
The advantage of the stochastic method is its flexibility, allowing to 
freely choose the momentum at the sink position 
as well as vary the timeslice of the current insertion.

Results from an initial test on a smaller gauge ensemble \cite{Alexandrou:2014pna} 
indicate that both methods show a compatible performance and give an approximately 
equal error for the same computational effort. Thus, for the following computations
the stochastic method will be used, since it is 
more flexible for studying 
larger momenta.

\subsection{Lattice setup}

All results shown in this work are computed on a 
$32^3 \times 64$ lattice from an ETMC (European Twisted Mass Collaboration) production ensemble \cite{Baron:2010bv},
with $N_f=2+1+1$ flavors of maximally twisted mass fermions, {\it i.e.} two degenerate light quarks and 
non-degenerate strange and charm quarks.
This ensemble has a bare coupling corresponding to $\beta=1.95$, 
which yields a lattice spacing of $a\approx 0.082$\,fm \cite{Alexandrou:2014sha}
and the twisted mass parameter $a\mu = 0.0055$, which corresponds to a pion mass of
$m_{PS} \approx 370$\,MeV.
Our present statistics to 
compute the matrix elements is 181 gauge configurations, each with
15 forward propagators at different source positions
and two stochastic propagators,
each propagator including both light (up and down) flavors, 
{\it i.e.} in total 5430 measurements.

\begin{figure}
	\centering
	\includegraphics[scale=0.8]{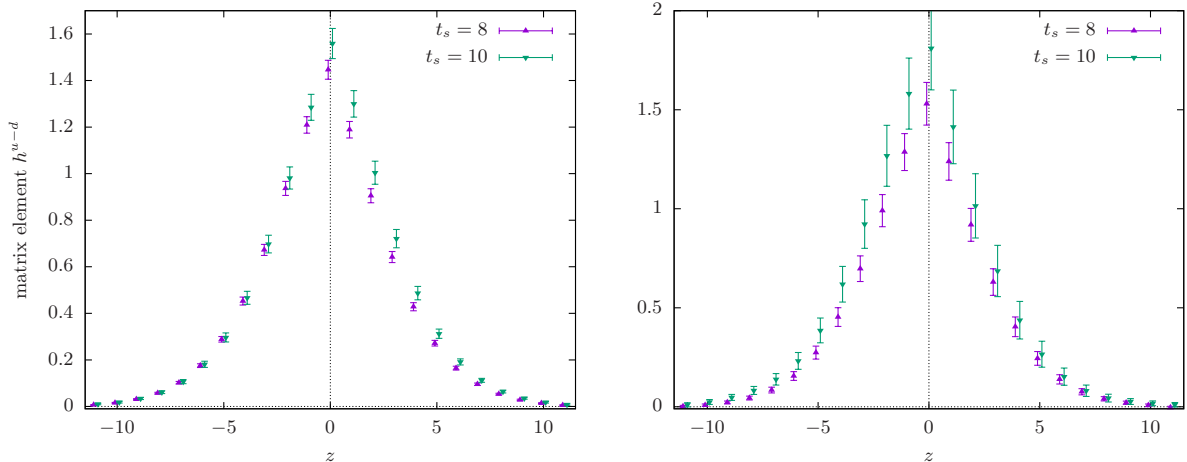}
	\caption{\label{FIG_PDF_ss}We show the results for the unrenormalized matrix elements for different source-sink separations $t_s$, {\bf left}: $P_3 = 2\pi/L$, {\bf right}: $P_3 = 4\pi/L$.}
\end{figure}

To examine the influence of excited states, the computation was done for
two different source-sink separations: $8a$ and $10a$.
From the comparison in Fig.\,\ref{FIG_PDF_ss}, it can be seen that
the results from both source-sink separations are visibly compatible within errors. 
It would require a significantly larger statistics to discriminate 
excited state effects, a task we want to address, however, in the future. 
Since here we perform an exploratory study, we will 
stick
to the small separation of $8a$ due to the significantly smaller noise
associated with it. 
This is especially advantageous 
for studying larger momenta, {\it e.g.} $P_3=6\pi/L$,
which has generically a bad signal-to-noise ratio.

\subsection{Lattice results}

\begin{figure}
	\centering
	\includegraphics[scale=0.8]{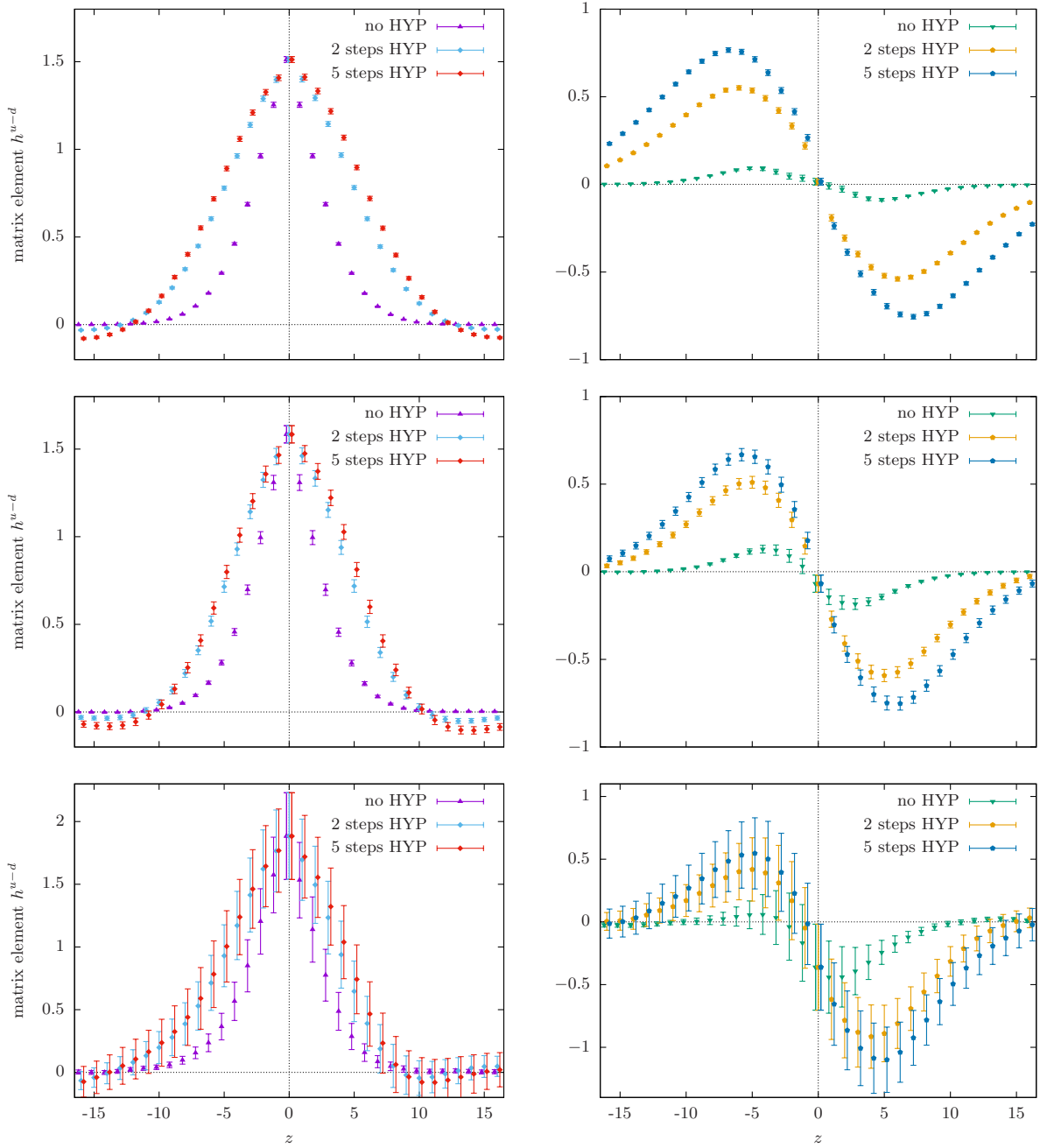}
	\caption{\label{FIG_COMPARE_HYP}Results for the unrenormalized matrix elements with different steps of HYP smearing, {\bf left}: real part, {\bf right}: imaginary part, {\bf from top to bottom}: $P_3=2\pi/L,4\pi/L,6\pi/L$.}
\end{figure}

For our current statistics, we were able to extract matrix elements for
$P_3 = 2\pi/L, 4\pi/L$ and $6\pi/L$.
In Ref.\,\cite{Lin:2014zya}, the authors applied HYP smearing \cite{Hasenfratz:2001hp} 
to the gauge links in the inserted operator.
This is a lattice technique, which is used to
smoothen the gauge links and 
is expected to bring the necessary
renormalization factors closer to the corresponding tree-level value.
More generally speaking, such kind of smearing will certainly influence 
the renormalization properties of the considered matrix elements. In order 
to obtain an estimate how renormalization could   
influence the results which will be presented here, 
we applied two and five steps of HYP smearing to
the operator and compare with the unsmeared results in Fig.\,\ref{FIG_COMPARE_HYP}.

Evidently, the effect of gauge link smearing changes the
value of the matrix elements, for both the real and the imaginary parts.
Note that 
the effect for the imaginary part is stronger than for the real part.
Also, the change from zero to two steps of smearing is more
significant than from two to five steps, which indicates a saturation 
of the smearing effect. 
We will therefore employ a maximum of five smearing steps in this work.
We note in passing that a 
decrease of the noise like in other gluonic 
quantities, {\it e.g.} as in \cite{Alexandrou:2013tfa}, cannot be observed 
when applying smearing.
A striking observation in Fig.\,\ref{FIG_COMPARE_HYP} is that while 
the real part is symmetric in $z$, the imaginary part is highly asymmetric.
This effect will play an important role when we discuss the quark and anti-quark 
distributions later on. 

Note that for a value of $z = 0$, 
the operator $\mathcal O$ in Eq.\,(\ref{eq:operator}) can be identified with 
the local vector current at $Q^2=0$. This operator is renormalized with 
the vector current renormalization constant $Z_V$, 
which, for this ensemble, is $Z_V=0.625(2)$ \cite{Alexandrou:2013joa}. 
After renormalization, the condition $F_1^{u-d}(Q^2=0)=1$ 
({\it cf.} \cite{Hagler:2009ni}) should hold. 
Indeed, we find $Z_V h^{u-d}(0)=1.18(22)$ for $P_3= 6\pi/L$ 
and $Z_V h^{u-d}(0)=0.99(3)$ for $P_3= 4\pi/L$
while the value for $P_3= 2\pi/L$, $Z_V h^{u-d}(0)=0.95(1)$, is a bit smaller,
which is probably due to excited state effects\footnote{
Using a larger source-sink separation of $10a$,  
we find $Z_V h^{u-d}(0)=0.98(4)$, {\it cf.} Fig.\,\ref{FIG_PDF_ss},
which is compatible with a value of one.
We attribute the larger error to the larger
source-sink separation and the fact that less measurements were used.}. 
For our final results, we will only use data obtained for 
$P_3=4\pi/L$ and $P_3=6\pi/L$. 

As can be seen in Fig.\,\ref{FIG_COMPARE_HYP}, when going to 
larger values of $P_3$, the signal-to-noise ratio rapidly worsens.
Thus, the calculation
of a further, larger momentum is not possible with our present statistics.

\section{Matching to quark distribution and nucleon mass corrections}

From the matrix elements $h^{u-d}(z,P_3)$, we calculate the quasidistributions and, 
after matching and nucleon mass corrections, the quark distributions
themselves. To this end, we first apply the Fourier transformation in Eq.\,(\ref{eq5})
to the nucleon matrix elements from $z=-L/2$ to $z=L/2$, after multiplying 
by the vector current renormalization constant $Z_V$. 
From this equation, it is clear that if the 
imaginary part of the matrix elements were zero, or very close to zero, 
there would be no difference between the positive and 
negative $x$ regions. In other words, there would 
be no difference between the quark and antiquark distributions, 
as antiquarks can be interpreted as quarks in the negative 
$x$ region, according to the crossing relation
$\overline{q}(x) = -q(-x)$. 

\begin{figure}
\centering
\includegraphics[scale=0.72]{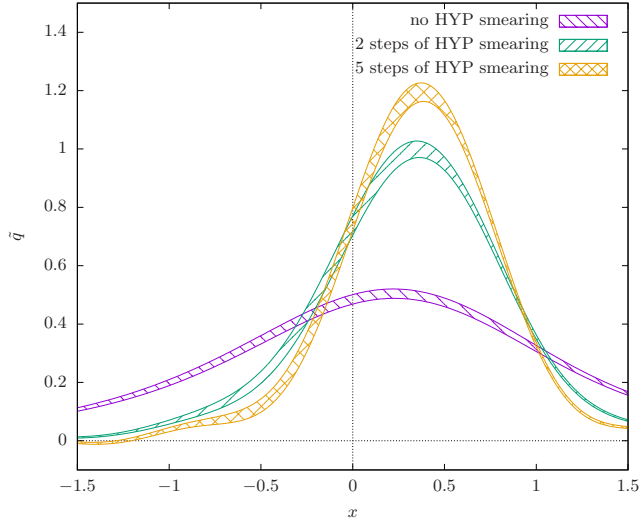}
\caption{\label{FIG_PDF_QTILDE}Comparison of results for $\tilde{q}$ obtained with five, two and no steps of HYP smearing, $P_3=4\pi/L$.}
\end{figure}

Fig.\,\ref{FIG_PDF_QTILDE} shows the complete quasidistribution for $P_3 = 4\pi/L$, after applying the 
Fourier transformation and taking the real and the imaginary parts of $h^{u-d}(z,P_3)$
into account. 
An asymmetry between  negative and positive $x$ values is clearly building up,  
which is more pronounced for higher levels of gauge link smearing, 
emphasizing the effect of HYP smearing on the renormalization of these quantities
Because after a proper renormalization 
the results with non-smeared and smeared gauge links have to agree within 
errors, the effect seen in Fig.\,\ref{FIG_PDF_QTILDE} clearly points to
the fact that renormalization will play an important role when  
looking at the quark distributions obtained from lattice calculations 
in the future. 

Having the quasidistribution $\tilde q(x)$, we can proceed
to extract the physical quark distribution $q(x)$ from $\tilde q(x)$, 
using Eq.\,(\ref{invq}) and then applying the nucleon mass corrections. 
To be consistent, the value of the momentum cutoff is chosen to be the same as 
the value of the lattice cutoff itself, that is, 
$\Lambda = 1/a \cong 2.5$\,GeV. For the renormalization scale $\mu_R$ we make the same choice. 
This is a somewhat {\it ad hoc}, but plausible choice. Once a proper renormalization has been 
carried out, the full equations for the running with $\mu_R$ will be obtained. 

As discussed in the Appendix, the integrals also have a cut-off at $x_c \sim \Lambda/P_3$, such that 
$\tilde q(x >x_c,\Lambda,P_3) =0$. The last input we need for our calculation is the bare coupling constant, 
for which we use the value corresponding to  
$\beta=1.95$ of our  lattice calculation. This leads to $\alpha_s = 6/(4\pi\beta) \approx 0.245$.

\begin{figure}
\centering
\includegraphics[scale=0.8]{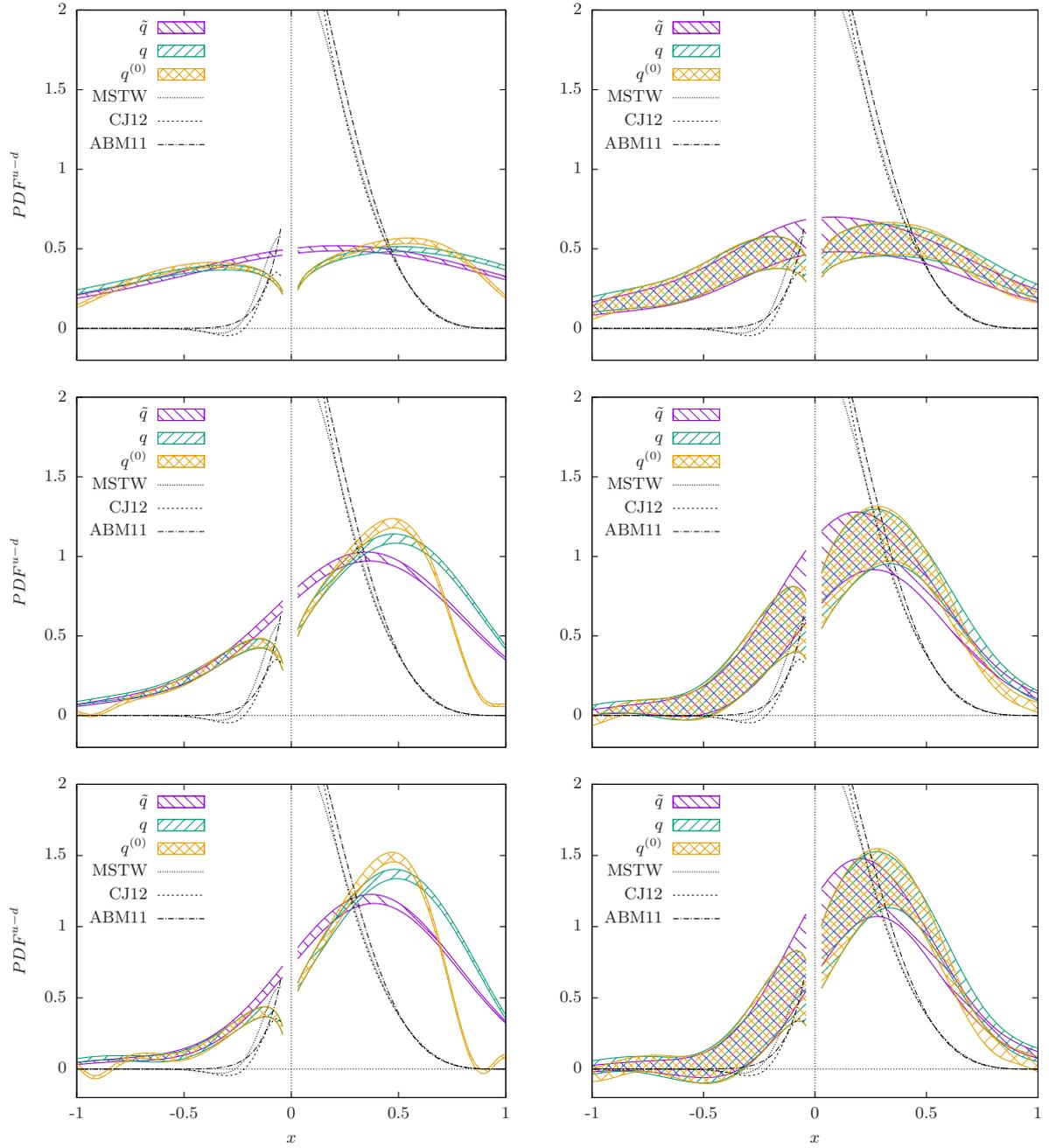}
\caption{\label{FIG_PDF_COMPLETE}The resulting quasidistribution $\tilde{q}$, 
PDF without subtracting the mass correction $q$, 
and final PDF $q^{(0)}$, {\bf left}: $P_3=4\pi/L$, {\bf right}: 
$P_3=6\pi/L$, {\bf from top to bottom}: 0 steps, 2 steps, 5 steps 
of HYP smearing, negative region: $\bar{q}(x)=-q(-x)$, comparison 
with phenomenological $u(x)-d(x)$ curves at $Q^2=6.25\,\text{MeV}^2$ 
(MSTW \cite{Martin:2009iq}, CJ12 \cite{Owens:2012bv}, ABM11 \cite{Alekhin:2012ig}). }
\end{figure}

We show our results in Fig.\,\ref{FIG_PDF_COMPLETE} 
for the case of 0, 2, and 5 steps of HYP smearing, 
for a nucleon with momentum $P_3 = 4\pi/L$ and $P_3 = 6\pi/L$, corresponding to 
0.98\,GeV and 1.47\,GeV, respectively. 
As anticipated, HYP smearing is essential in providing the 
required asymmetry between the quark and antiquark distributions. We note 
that two steps of smearing are already practically 
sufficient to show the effect of the asymmetry. 

As the nucleon momentum increases, the peak of the $u(x)-d(x)$ distribution 
moves to smaller values of $x$, as it should, 
while for $\overline d(x) - \overline u(x)$ it gets closer to 0 for most 
of the $x$ region, but shows an increase in the small $x$ region. This behavior 
is in qualitative agreement with the behavior of the antiquark distributions
as extracted from phenomenological analyses \cite{Martin:2009iq,Owens:2012bv,Alekhin:2012ig}. 
The nucleon mass corrections 
lead to a decrease of the distributions in the large $x$ region. 
This is again in full accordance with our expectation from phenomenology  
and asserts that the nucleon mass corrections are essential to restore 
the energy-momentum relations, thus ensuring 
the partonic interpretation of the distributions. In addition, with increasing 
nucleon momentum 
the mass corrections become less and less important, as expected from Eq.\,(\ref{eq11}).
Finally, the slight oscillatory behavior in the large $x$ region is a 
result of performing the Fourier transformation over a finite extension only,
in our case the integration is from $-L/2$ to $+L/2$. 
Because the nucleon mass corrections also make a shift of the distributions 
from larger to smaller values of $x$, the oscillatory behavior 
is more pronounced after Eq.\, (\ref{eq11}) is applied, as the oscillations are more noticeable 
in the quasidistributions in the region $x>1$. 
Increasing the value of $P_3$ is similar to extending the 
bounds of integration and thus reduces the oscillations. On this same line, if we had used $\pm 1$
as the limits of of integration in the matching, as in Eq.\,(\ref{eq9}) and thus not taking into account the $x>\xi$ region in the last two terms of Eq. \,(\ref{invq}), the
oscillations would be slightly more pronounced for the case of $P_3 =4\pi/L$. For $P_3=6\pi/L$ there would be
no real difference to the results presented in  Fig.\,\ref{FIG_PDF_COMPLETE}.

\begin{figure}
\centering
\includegraphics[scale=0.8]{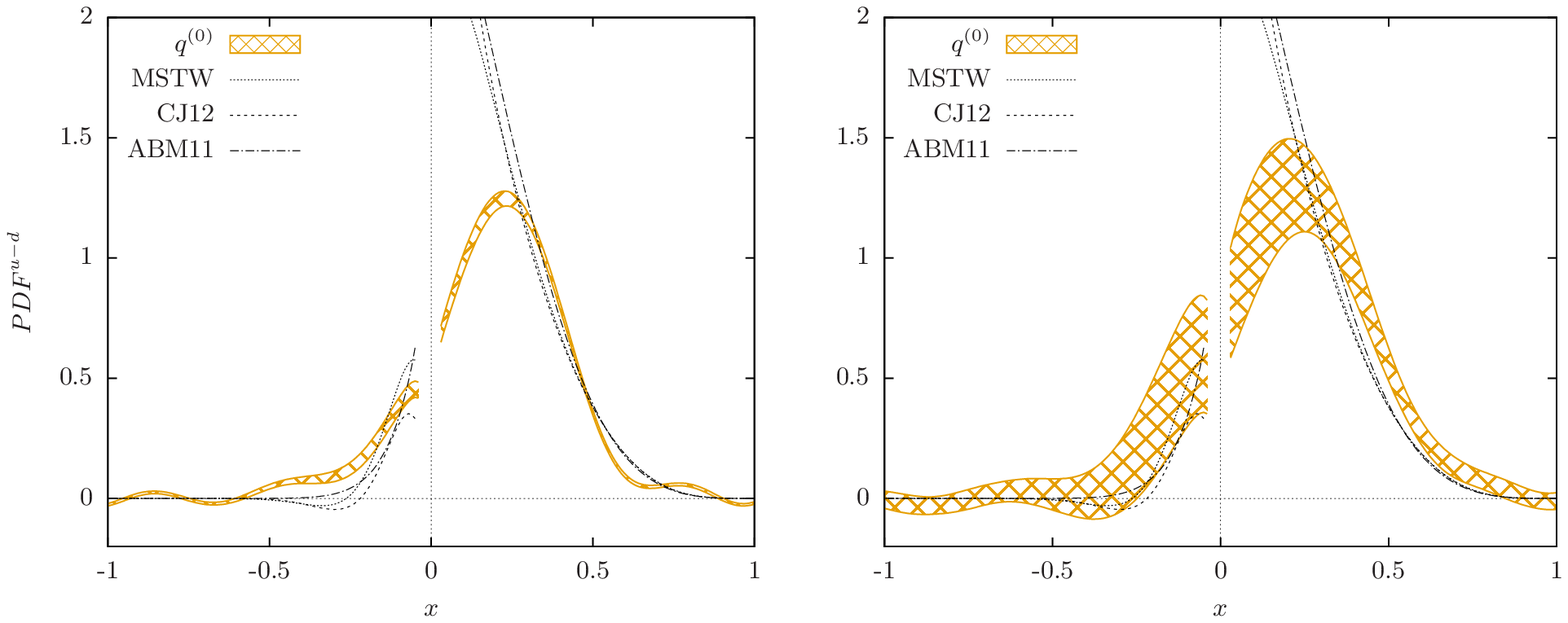}
\caption{\label{FIG_PDF_MIXED}Results from an hypothetical mixed momentum analysis using different values of the momentum in the computation of the lattice matrix element ({\bf left}: $P_3=4\pi/L$, {\bf right}: $P_3=6\pi/L$) than in the Fourier transformation, matching and mass corrections ($P_3=8\pi/L$) with 5 steps of HYP smearing.}
\end{figure}

Although we find that the shape of the quark distributions
resembles those of the 
phenomenological parametrizations of $u(x)-d(x)$, with 
2 or 5 steps of HYP smearing, we do not find an agreement
on the quantitative level. Note, however, that there is a clear tendency 
to approach the phenomenological parametrizations when $P_3$ is 
increased. 
Motivated by this observation, we made an 
exploratory study where we use the matrix elements 
calculated with $P_3 = 4\pi/L$ and $P_3 = 6\pi/L$, but perform 
the Fourier transformation in Eq.\,(\ref{eq5}),
as well as the matching and the nucleon mass corrections, with $P_3 = 8\pi/L$.
We will refer to this particular setup as the mixed momentum setup. 
The resulting distributions are shown in Fig.\,\ref{FIG_PDF_MIXED}.

It needs to be stressed that this exercise is, of course, only hypothetical,
and using this setup can lead to the loss of normalization of the
parton distributions, opposed to the results in Fig.\,\ref{FIG_PDF_COMPLETE}, where we found a
good agreement with a normalization of one.  
Nevertheless, 
the agreement with the phenomenological parametrizations of the distributions at the
intermediate and large $x$ regions is really encouraging. This indicates
that by employing an only moderately larger value of $P_3$ than the ones used
here, we could obtain even a quantitative agreement to the parametrizations in certain regions of $x$.
This concerns in particular the large $x$ region, where 
increasing values of $P_3$  
tend to bring the resulting distribution down. 
In the small (and positive) $x$ region, on the other hand, 
it seems that increasing the nucleon momentum is not sufficient 
to produce a rise of the distribution. 
This may be related to the fact that there is a limitation in  
the present calculation in the small $x$ region due to the presence
of the infrared, $1/L$, and ultra-violet, $1/a$ cut-off regulators
on a finite lattice. Thus, this limitation will be overcome when 
larger lattices and smaller values of the lattice spacing become
available.
Furthermore, we stress that the here obtained results are at
only one, non-physical value of the pion mass
and the shape of distribution might as well depend on
the quark mass.
In any case, a more definite statement can only be made    
after we have access to the matrix elements 
for $P_3 = 8\pi/L$. This is not possible with our present statistics. 
However, we are in the process of generating a substantially higher statistics. 
This will allow us to extrapolate the data for $P_3=2\pi/L$, $P_3=4\pi/L$ and $P_3=6\pi/L$
to obtain the quasidistribution at $P_3=8\pi/L$. Although we do not expect 
a big difference to the situation of the hypothetical mixed setup shown 
in Fig.\,\ref{FIG_PDF_MIXED}, 
a full analysis with real data is, of course, mandatory and will be presented in a forthcoming work. 

\section{Conclusions and outlook}

In this work, we have presented our first exploratory study
of the approach developed in Ref.\,\cite{Ji:2013dva} for the calculation
of the $x$ dependence of quark distributions directly on the lattice,
employing the twisted mass formulation of lattice QCD. 
The study presented here, together with the work of Ref.\,\cite{Lin:2014zya}, 
constitute the first two attempts to implement 
the approach of Ref.\,\cite{Ji:2013dva} in realistic lattice QCD calculations. 
Our results, represented in Fig.\,\ref{FIG_PDF_COMPLETE},
are comparable to those of Fig.\,2 of \cite{Lin:2014zya} 
and we see that the two calculations give similar results 
for the case of 2 steps of HYP smearing. 
Yet it seems that in our case, for $P_3 = 6\pi/L$, 
the shift of the peak of the quark distributions 
towards the small $x$ region is bigger. 
Also, as we increase the number of steps of HYP smearing, 
the position of the peaks is basically unchanged, 
but they are more pronounced. 
On the other hand, the large $x$ region tends to be smaller. 
It is very reassuring to see that 
both effects bring our results closer to the phenomenological parametrizations. 
Beyond these effects, 
HYP smearing is fundamental to generate a sizeable 
(asymmetric in $z$) imaginary part in the matrix elements.
This result 
generates automatically an asymmetry between the quark and the antiquark 
distributions, a highly non-trivial result.

The outcome of our {\it ab initio} lattice QCD calculation with a small and 
positive ${\overline{d}(x)-\overline{u}(x)}$ is in a 
very good qualitative agreement with phenomenological parametrizations. 
In a hypothetical exercise where we use a larger momentum of $P_3=8\pi/L$ in the 
Fourier transformation than we actually have in our lattice QCD calculation, 
we observe a better qualitative behavior as compared to what is expected phenomenologically, 
as is shown in Fig.\,\ref{FIG_PDF_MIXED}. 
Moreover, it is clear from both Figs.\,\ref{FIG_PDF_COMPLETE} 
and \ref{FIG_PDF_MIXED} that increasing the momentum 
implies only marginal corrections to the quasidistributions, 
the corrections for the case $P_3 = 6\pi/L$ being restricted 
from intermediate to small $x$ regions only. 

In summary, we have presented our first effort 
to explore the potential 
to calculate quark distributions directly within the lattice QCD formulation. 
Although there are clearly shortcomings, such as not being able
to reach large momenta and the lack of renormalization, our 
results are promising. In particular, our study of the quark distribution
in the mixed momentum setup indicates that only moderately larger momenta 
than used here may be sufficient to reach a quantitative agreement with 
phenomenological parametrizations in the large $x$ region. 
We are presently increasing our statistics significantly, which will allow 
us to obtain data with such larger momenta. 
In addition, we are testing different approaches to perform the necessary 
renormalization of the matrix elements entering the calculation 
of the quasidistributions. 
Finally, applying this method to gluon configurations generated directly at 
the physical value of the pion mass may open the exciting possibility 
to address quark distributions and therefore 
unravel the structure of hadron from first principle QCD calculations. 

\section*{Acknowledgments}
We thank our fellow members of ETMC for their constant collaboration. In particular helpful
discussions with G.C. Rossi are gratefully acknowledged. 

We are grateful to the John von Neumann Institute for Computing (NIC),
the J{\"u}lich Supercomputing Center and the DESY Zeuthen Computing
Center for their computing resources and support.
  
This work has been supported in part by the 
DFG Sonderforschungsbereich/Transregio SFB/TR9
and by the Cyprus Research Promotion
Foundation through the  Project Cy-Tera 
(NEA Y$\Pi$O$\Delta$OMH/$\Sigma$TPATH/0308/31) co-financed by the European Regional
Development Fund. 
KC has been supported in part by the Helmholtz International Center for FAIR within the
framework of the LOEWE program launched by the State of Hessen.
VD was supported by the Danish National Research Foundation 
DNRF:90 grant and by a Lundbeck Foundation Fellowship grant.
FS was supported by CNPq contract number 249168/2013-8.

\pagebreak

\appendix

\section*{Appendix}

The wave function and vertex corrections in Eq.\,(\ref{eq9}) 
were calculated in Ref.\,\cite{Xiong:2013bka}. The vertex corrections are given by:

\begin{equation}
\label{Z1}
\frac{Z^{(1)}(\xi)}{C_F} = \left(\frac{1+\xi^2}{1-\xi}\right)\ln\frac{\xi}{\xi-1} + 1 +
\frac{1}{(1-\xi)^2}\frac{\Lambda}{P_3}
\end{equation} 
for $\xi>1$,
\begin{equation}
\label{Z2}
\frac{Z^{(1)}(\xi)}{C_F} = \left(\frac{1+\xi^2}{1-\xi}\right)\ln\frac{(P_3)^2}{\mu_R^2} 
+ \left(\frac{1+\xi^2}{1-\xi}\right) \ln 4\xi(1-\xi)
-\frac{2\xi}{1-\xi}
+ 1 +
\frac{1}{(1-\xi)^2}\frac{\Lambda}{P_3}
\end{equation} 
for $0<\xi<1$,
\begin{equation}
\label{Z3}
\frac{Z^{(1)}(\xi)}{C_F} = \left(\frac{1+\xi^2}{1-\xi}\right)\ln\frac{\xi-1}{\xi} - 1 +
\frac{1}{(1-\xi)^2}\frac{\Lambda}{P_3}
\end{equation} 
for $\xi<0$. The wave function corrections are given by:

\begin{equation}
\label{dZ}
\delta Z^{(1)} =  C_F \int_{-\infty}^\infty d\xi \, \delta Z^{(1)}(\xi),
\end{equation} 
where
\begin{equation}
\label{dZ1}
\delta Z^{(1)}(\xi) = -\left(\frac{1+\xi^2}{1-\xi}\right)\ln\frac{\xi}{\xi-1} - 1 -
\frac{1}{(1-\xi)^2}\frac{\Lambda}{P_3}
\end{equation} 
for $\xi>1$,
\begin{equation}
\label{dZ2}
\delta Z^{(1)}(\xi) = -\left(\frac{1+\xi^2}{1-\xi}\right)\ln\frac{(P_3)^2}{\mu_R^2} 
- \left(\frac{1+\xi^2}{1-\xi}\right) \ln 4\xi(1-\xi)
+\frac{2\xi(2\xi-1)}{1-\xi} 
+ 1 -
\frac{1}{(1-\xi)^2}\frac{\Lambda}{P_3}
\end{equation} 
for $0<\xi<1$,
\begin{equation}
\label{dZ3}
\delta Z^{(1)}(\xi) = -\left(\frac{1+\xi^2}{1-\xi}\right)\ln\frac{\xi-1}{\xi} + 1 -
\frac{1}{(1-\xi)^2}\frac{\Lambda}{P_3}
\end{equation} 
for $\xi<0$.

In the actual calculation, we make a change of variables in the integral 
term containing $\tilde q(y,\Lambda,P_3)$  of Eq.\,(\ref{eq9}), and also set the threshold above which
the quasidistribution is zero. We call this value $x_c$, which is of order of $\Lambda/P_3$. When we inverted Eq.\,(\ref{eq8}),
we kept the limits of integration from -1 to +1, which is the region where the quark distributions are defined
and where factorization holds. In practice, we will integrate from $-x_c$ to $+x_c$, the reason being that 
$\tilde{q}(x>1) \neq 0$ and contributions from this region should be taken into account. As we increase 
the value of $P_3$, however, the closer we get to the physical distribution and as a result $\tilde{q}(x>1) \sim 0$.
We also break the integral containing $\tilde q$   
into two terms, with the limits from $- x_c$ to $-|x|/x_c$ 
and from $+|x|/x_c$ to $+x_c$. We then make a change of
variables, $\xi = x/y$, and Eq.\,(\ref{eq9}) is rewritten as:

\begin{eqnarray}
\label{invq}
 q(x, \mu_R) &=& \tilde{q} (x,\Lambda,P_3) - \frac{\alpha_s}{2\pi} \tilde{q} (x,\Lambda,P_3)  \delta Z^{(1)} \left( \frac{\mu_R}{P_3}, \frac{\Lambda}{P_3}\right)  \nonumber\\*
& & - \frac{\alpha_s}{2\pi} \int_{-x_c}^{-|x|/ x_c} Z^{(1)} \left( \xi, \frac{\mu_R}{P_3}, \frac{\Lambda}{P_3} \right) \tilde{q} \left( \frac{x}{\xi} ,\Lambda,P_3 \right) \frac{d\xi}{|\xi |} \nonumber\\*
& & - \frac{\alpha_s}{2\pi} \int_{+|x|/ x_c}^{+x_c} Z^{(1)} \left( \xi, \frac{\mu_R}{P_3}, \frac{\Lambda}{P_3} \right) \tilde{q} \left( \frac{x}{\xi} ,\Lambda,P_3 \right) \frac{d\xi}{|\xi |} +  \mathcal{O}(\alpha_s^2). 
\end{eqnarray}
The integrals contain both single and double poles 
at $\xi = 1$. It can be shown that the single pole terms cancel 
between Eqs.\,(\ref{Z1})-(\ref{Z2}) and (\ref{dZ1})-(\ref{dZ2}), {\it e.g.} the single
pole in the third term on the l.h.s of Eq.\,(\ref{Z2}) is cancelled by the third term on the l.h.s of Eq.\,(\ref{dZ2}). The double poles are first 
reduced to a single pole by a similar cancellation when 
combining the vertex and wave function corrections, as in the single pole case, and the 
remaining pole is taken care of by using the Cauchy's principal value prescription. 
The remaining expression is finite, 
with the exception  that the integral of $\delta Z^{(1)}(\xi)$ 
is divergent as $\xi \rightarrow \pm \infty$. The divergent term is:

\begin{equation}
\tilde{q}(x)\frac{3}{2}\ln(x_c^2-1),
\label{UVdiv}
\end{equation}
where we have set $x_c$ as the upper and lower limit of the integrals 
of $(\ref{dZ1})$ and $(\ref{dZ3})$, respectively. The same limits of integration, both when
integrating $Z^{(1)}$ and $\delta{Z}^{(1)}$ are necessary
in order to maintain the quark number conservation. 
Notice that this log divergent term is the usual UV divergence 
present in the wave function. 

\pagebreak

\bibliographystyle{utphys}
\bibliography{./references.bib}

\end{document}